\documentclass[letter]{aa}  

\usepackage{graphicx}
\usepackage{natbib}
\usepackage{txfonts}
\begin{document} 

\title{Chromospheric impact of an exploding solar granule}

   \author{C.E.~Fischer
          \inst{1}
          \and
           N.~Bello Gonz\'alez
           \inst{1}
           \and
           R.~Rezaei
           \inst{2,3}
}

   \institute{Kiepenheuer Institut f\"{u}r Sonnenphysik, Sch\"{o}neckstrasse 6, 79104 Freiburg, Germany
                     \and
             Instituto de Astrof\'{i}sica de Canarias, Avda V\'{i}a L\'{a}ctea S/N, La Laguna 38200, Tenerife, Spain
        \and
             Departamento de Astrof\'isica, Universidad de La Laguna, 38205 La Laguna (Tenerife), Spain
}


 
  \abstract
   {Observations of multi-wavelength and therefore height-dependent information following events throughout the solar atmosphere and unambiguously assigning a relation between these rapidly evolving layers are rare and difficult to obtain. Yet, they are crucial for our understanding of the physical processes that couple the different regimes in the solar atmosphere. }  
   {We characterize the exploding granule event with simultaneous observations of  $Hinode$  spectroplarimetric data in the solar photosphere and $Hinode$ broadband Ca\,{\sc ii} H images combined with Interface Region Imaging Spectrograph ($IRIS$) slit spectra. We follow the evolution of an exploding granule and its connectivity throughout the atmosphere and analyze the dynamics of a magnetic element that has been affected by the abnormal granule. }
  {In addition to magnetic flux maps we use a local correlation tracking method to infer the horizontal velocity flows in the photosphere and apply a wavelet analysis on several $IRIS$ chromospheric emission features such as Mg\,{\sc ii}\,k2v and Mg\,{\sc ii}\,k3 to detect oscillatory phenomena indicating wave propagation.}
   {During the vigorous expansion of the abnormal granule we detect radially outward horizontal flows, causing, together with the horizontal flows from the surrounding granules, the magnetic elements in the bordering intergranular lanes to be squeezed and elongated. In reaction to the squeezing, we detect a chromospheric intensity and velocity oscillation pulse which we identify as an upward traveling hot shock front propagating clearly through the $IRIS$ spectral line diagnostics of Mg\,{\sc ii}\,h$\&$k.}
   {Exploding granules can trigger upward-propagating shock fronts that dissipate in the chromosphere.}

\maketitle
%

\section{Introduction}\label{sec:intro}
Exploding granules are a readily seen event in the granulation pattern in the photosphere estimated to cover $2.5~\%$ of the solar surface at any given time  \citep{1986A&A...161...31N}. They are bright granules that show a rapid expansion with expansion rates of $1.7$ to $3.2$~km\,s$^{-1}$ and reach diameters as large as  $5.5$\arcsec. Further, they show a dark core in the center, giving them the overall appearance of a bright ring before splitting into smaller granules \citep[see e.g.,][]{{1986A&A...161...31N},{1995ApJ...443..863R},{1999ApJ...527..405H}}. Their temporal evolution in intensity and velocity are well described in \cite{2001A&A...368..652R}. 
More recently, \cite{2012A&A...537A..21P} analyzed high-resolution spectropolarimetric {\em IMaX/SUNRISE} data and found that the evolution of several mesogranular-sized exploding granules is linked to magnetic flux emergence. \\
In the events described by \cite{1999ApJ...527..405H} it becomes clear that the individual evolution of each exploding granule is shaped by the surrounding material and the conditions present when the granule expands. Often material is observed to be jammed at the borders of the exploding granule which cannot expand freely. 
With the high-resolution {\em IRIS} spectrograph recording the Mg\,{\sc ii}\,k$\&$h 2796\,\AA\, spectral region, we are able to link the evolution of a magnetic element trapped by an exploding granule and the neighboring granules with the formation of a shock wave in the chromospheric layer. \vspace{-0.4cm}


\section{Observations}\label{sec:obs}
 \begin{figure*}
   \sidecaption
  \includegraphics[width=12cm]{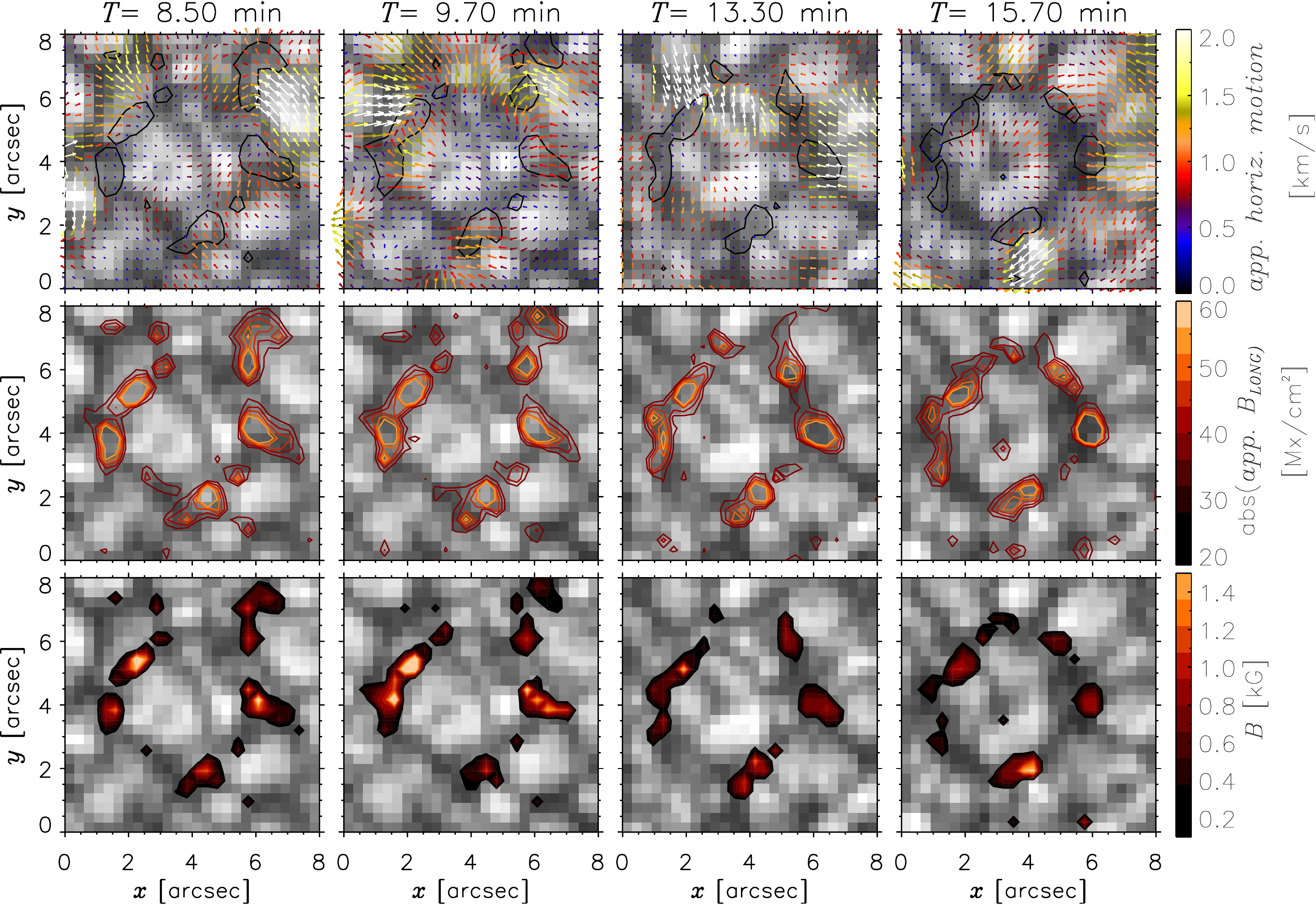}
     \caption{{\em Rows, Top to Bottom:} The first row displays the {\em Hinode} SP continuum images with the 2D field of the apparent horizontal velocities calculated by the Local Correlation Tracking (LCT) method shown with colored arrows. Black contours denote B$_{long}^{app}$ at $\pm\,30$\,Mx\,cm$^{-2}$ and the times above the panels refer to the {\em IRIS} time line. In the second row the continuum images are now overplotted by contours of the absolute of B$_{long}^{app}$ magnetic flux density from absolute $20$\,Mx\,cm$^{-2}$ to $60$\,Mx\,cm$^{-2}$ in steps of $10$\,Mx\,cm$^{-2}$. The last row shows the same continuum images now with filled contours of the magnetic field strength obtained from MERLIN inversions obtained  with the default inversion parameters from {http://www2.hao.ucar.edu/csac}.}
      \label{fdev}
        \end{figure*}

We analyzed data taken simultaneously with the {\em Hinode} \citep{2008SoPh..249..167T} and {\em IRIS} satellites \citep{2014SoPh..289.2733D}  on 28 April 2014 of a quiet Sun area at a heliocentric angle of $\mu=0.98$. The {\em Hinode} data consisted of binned {\em Hinode} Spectropolarimeter (SP) full Stokes 6302\,\AA\, slit spectra (spatial sampling of $\sim0.32$\arcsec) covering a region of  about $9\times81$~arcsec$^2$, spectral sampling of 21.5\,m\AA, and a cadence of $\sim70$\,s. The level 0 data were calibrated using the SolarSoft {\em sp$\_$prep.pro} routine~\citep{2013SoPh..283..601L}. The calibration routine also provides a file {\em $stksimg.sav$} with maps for the apparent magnetic flux densities (B$_{long}^{app}$ and B$_{trans}^{app}$), as well as continuum maps constructed from the continuum level of the Stokes $I$ spectra \citep{2008ApJ...672.1237L}.  Ca\,{\sc ii}\,H images taken with the {\em  Hinode} Broadband Filter Imager (BFI) at a cadence of $31$\,s, exposure time of $0.3$\,s, a field-of-view (FOV) of $38\times89$~arcsec$^2$, and a spatial sampling $\sim0.109$\arcsec were processed using the {\em fg$\_$prep.pro} routine.\\
The {\em IRIS} satellite provided a slitjaw image time series in the $2796$\,\AA\,band at a spatial sampling of $0.166$\arcsec, a FOV of  $62\,\times\,66$~arcsec$^2$ and a cadence of $19$\,s. Slit spectra were obtained in a two-step raster ({\em IRIS}-program: medium sparse 2-step raster) with a cadence of $19$\,s for each of the two slit positions recording FUV and NUV spectra in several wavelengths of which we show results in the region of Mg\,{\sc ii}\,h$\&$k at 2796\,\AA\,.
The spatial sampling along the slit was $0.166$\arcsec and the step size of the $0.33$\arcsec wide slit was $1$\arcsec. We obtain level $2$ data by running the {\em iris$\_$make$\_$lev2.pro} code. The code performs dark and flat calibration as well as an initial alignment between the slitjaw images and the spectra.

The data from both instruments were aligned using strong network magnetic elements appearing as brightenings in the {\em Hinode} Ca\,{\sc ii}\,H images and {\em IRIS} slitjaw images and rebinned to the {\em IRIS} slitjaw pixel size of $0.166$\arcsec. We created data cubes at the cadence of the {\em IRIS} slitjaw images by selecting for all instruments for each {\em IRIS} slitjaw timestamp the images closest to this timestamp. We did not perform an interpolation in time, which leads to repetition in the images for the cubes with a lower cadence (as seen in the online movies). The maximum error $\delta$T in assigning the frames in reference to the {\em IRIS} slitjaw images (fastest cadence) can arise for the {\em Hinode} SP and is of the order of one minute.

\section{Results and discussion}
\label{sec:res}
In the first panel of the movie attached to Fig.~\ref{Time_evolv_f} one can follow the temporal evolution of the exploding granule as seen in the constructed {\em Hinode} SP continuum maps. The time $T=0\,\rm min$ refers to the beginning of the {\em IRIS} sequence starting at 11:39:36\,UT. A small bright granule (area of $1.35$\,arcsec$^{2}$) is seen in the {\em Hinode} SP continuum map cutout located at $x=5\arcsec$ and $y=4.5\arcsec$ just two and a half minutes into the co-observing sequence. The granule explodes by expanding with a rate of about $\sim0.77$ arcsec$^{2}$ per minute, and then develops a dark core before splitting into several fragments, finally reaching a ring shaped form with an overall size of about $14$\,arcsec$^{2}$ at $T=19\,\rm min$. The morphological evolution of our exploding granule in the photosphere is consistent with \citet{1999ApJ...527..405H}.\\
There is a group of mixed polarity strong magnetic elements (\textgreater\,$500$\,G) that occupies the intergranular lanes in the vicinity of the exploding granule. After the exploding phase they are arranged in a circle around the granule, outlining its border and tracing the same ring-like pattern as seen in the {\em Hinode} SP continuum maps. The magnetic elements appear as bright points in the Ca\,{\sc ii}\,H images (third panel in the movie) which again line up as a chain encompassing the exploded granule at the end of the sequence. The bright points are also visible in the {\em IRIS} $2796$\,\AA\,band slitjaw images (last panel), as is the location of the slit at the edge of the exploding granule. By this point the interior features several dark lanes and after the maximum size is reached (at about $19$ - $20$\,min) the shape looses coherency and, after a few minutes of granular evolution, cannot be traced anymore. \\
\begin{figure*}
\centering  
\resizebox{16cm}{!}{ \includegraphics{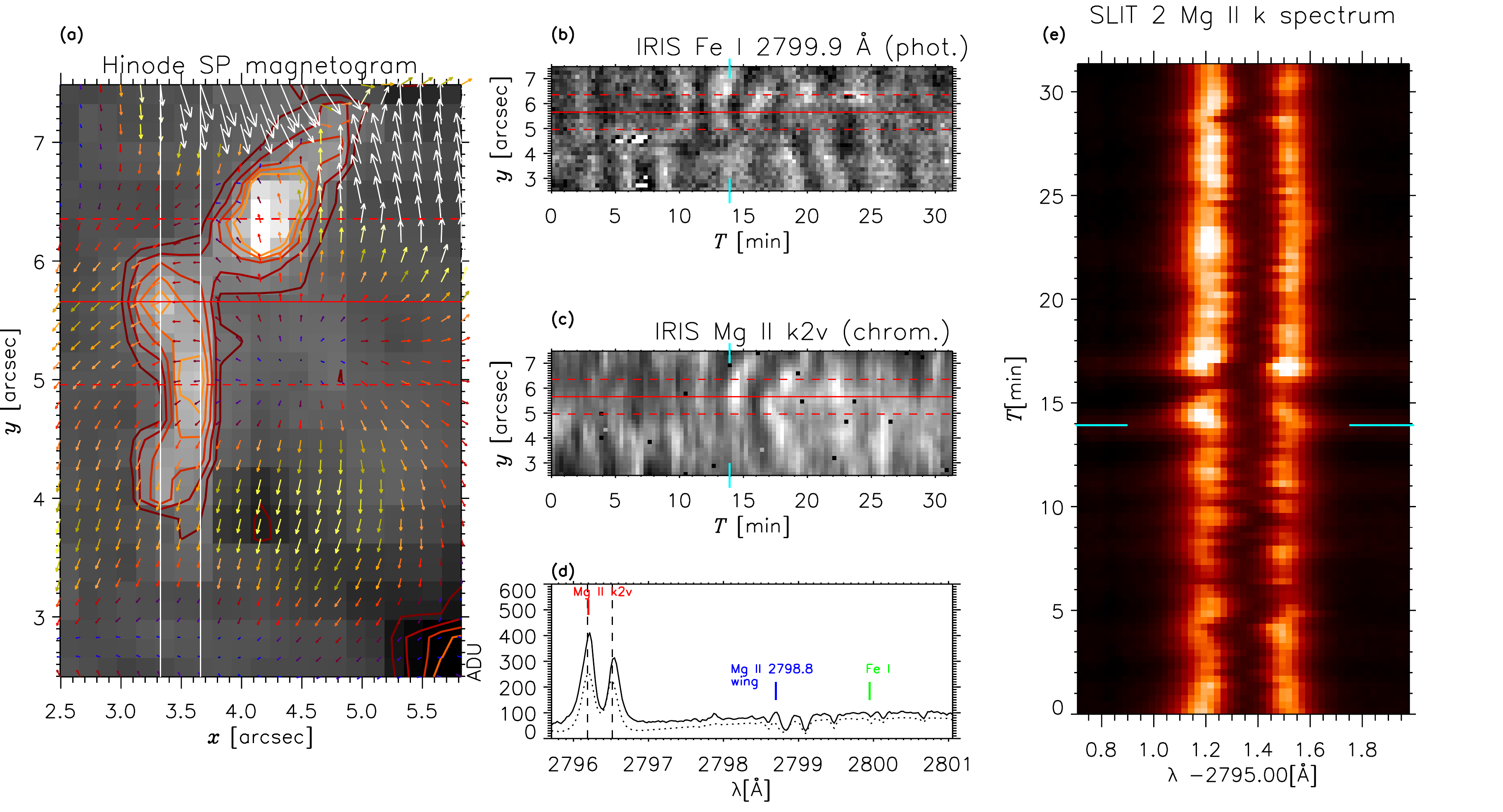}}
\caption{{\em Left to Right:} Panel (a) shows the {\em Hinode} SP B$_{long}^{app}$ map now rebinned to the {\em IRIS} slitjaw scale with the same contours as in the second row of Fig.~\ref{fdev} and the horizontal flows indicated with colored arrows as in Fig.~\ref{fdev}. The two vertical white lines delimit the position of the {\em IRIS} slit in use. Two dashed red horizontal lines and a solid red line mark a region of interest. Panels (b) and (c) are time-space images derived from spectra recorded by the second {\em IRIS} slit. Panel (b) shows the line-core velocity of the Fe\,{\sc i}\,$2799.9$\,\AA\, line ($\pm\,1.9$ km\,s$^{-1}$,  where down flows are positive)  retrieved with the MOSiC code~\citep{2017arXiv170104421R} and in panel (c) the intensity of the Mg\,{\sc ii}\,k2v peak (in data units) obtained by the ${iris\_get\_mg\_features\_lev2.pro}$ code is plotted. The red horizontal lines indicate the same area as in panel (a). The short vertical lines in cyan indicate the time of the map in panel (a), and panel (d) shows a spectrum at that time (solid) and average quiet Sun spectrum (dotted).  The dashed vertical lines denote the average position of the Mg\,{\sc ii}\,k2 peaks.
%
The right panel demonstrates the temporal evolution of the {\em IRIS} slit spectra at the slit location marked by the solid red horizontal line in panels (a) to (c) with the time of panel (a) indicated again with lines in cyan.   \label{specev}}
\end{figure*}
We use the Fourier Local-Correlation-Tracking (FLCT) code  \citep{Fisher:2007jf}  to obtain two-dimensional (2D) velocity fields inferred from the {\em Hinode} SOT SP continuum images. These continuum images were constructed using the continuum intensity in Stokes I for each slit position and building up the map with a cadence for the entire FOV of about 70\,s. We set the full width at half maximum (FWHM) of the Gaussian used to weigh the pixels for the cross-correlation to four pixels ($1.2\,\arcsec$). 
Running the code on pairs of two subsequent images of the time series, we obtain the apparent horizontal velocities for structures on a granular scale.  The obtained results displayed in Fig.~\ref{fdev} show velocities greater than $1.5$\,km\,s$^{-1}$ from the outer granules impinging on the exploding granule. We note that these velocities are only apparent horizontal velocities and are highly dependent on the structure scale of the image and the choice of FWHM of the employed Gaussian weighing function. Especially in a bright expanding larger region, the velocities might be underestimated in the interior due to no apparent structure changes.
At $T=9.7$\,min, the magnetic elements at the edge of the exploding granule at around  $x=2\arcsec$ and $y=5\arcsec$ experience strong horizontal velocities from both sides; the interior of the exploding granule and the surrounding granules. This gives the impression of the magnetic elements being squeezed by the opposing flows. This is confirmed by the magnetic flux density evolution. The contours of the absolute B$_{long}^{app}$  maps show a stretching out of the magnetic elements, leading to a more elongated shape (compare, for example, panel one at $T=8.5$\,min and panel three at $T=13.3$\,min in Fig.~\ref{fdev} and separation of the magnetic flux elements. This is accompanied by a decrease in the total magnetic field strength as seen in the last row of Fig.~\ref{fdev} and is therefore not due to a change of the inclination angle tilting the magnetic field.\\
Figure~\ref{specev} shows the time development of the (high) photospheric line-of-sight velocity from fits to the Fe\,{\sc i}\,2799.9\,\AA\, line and the  {\em IRIS} Mg\,{\sc ii}\,k2v peak intensity (a good indicator of the chromospheric temperature according to \cite{2013ApJ...778..143P}) in a spatial section along the slit where blacked-out pixels indicate locations where the feature finding failed.
In both diagnostics, a similar oscillatory pattern is seen, although out of phase, and starting at around $T=10$\,min for the photosphere and around 12\,min for the chromosphere lasting for a few minutes and located at the area $\pm0.7$\arcsec of the magnetic element location observed in the {\em Hinode} SP magnetic flux density maps. The oscillatory pattern appears one to two minutes after the previously described squeezing of the magnetic elements. As seen in the  sample spectra in panel (d) we observe the wing of the Mg\,{\sc ii}\,2798.8\,\AA\, subordinate line going into emission. According to \cite{2015ApJ...806...14P} this is a sign of heated material in the low chromosphere with cooler material above; a rare quiet Sun event  (1\% of pixels) also observed in simulations \citep[see also ][]{2015ApJ...809L..30C}. \cite{2013A&A...560A..50S} have observed similarly rare emission in the Fe\,{\sc i}\,3969\,\AA\, line out of phase with the simultaneously observed Ca\,{\sc ii}\,H2v intensity as a marker of chromospheric dynamics. \\
Panel (e) in Fig.~\ref{specev} shows the Mg\,{\sc ii}\,k spectra in time for the slit position marked with a red solid line in the previous panels. After an initial amplitude increase of the Mg\,{\sc ii}\,k2v peak at about $11.5$\,min, a dramatic sudden amplitude increase occurs at around $14$\,min. At $T=17$\,min the red emission peak Mg\,{\sc ii}\,k2r brightens followed again by an increase of the Mg\,{\sc ii}\,k2v peak.  
We find a switching of the amplitude increases in the two k2 peaks (see spectrum between $T=14.52$\,min and $T=16.78$\,min). The ratio of the Mg\,{\sc ii}\,k2v to Mg\,{\sc ii}\,k2r peaks is a good indicator of the chromospheric velocity above the peak formation height  \citep{2013ApJ...778..143P} with a higher peak in the blue emission peak corresponding to down flows. The movie attached to Fig.~\ref{Spec_evolv_f} shows the detailed temporal evolution of the spectra with an increase of the near continuum compared to the average profile clearly seen, for example, at $T=16.78$\,min with the entire spectrum being shifted to higher intensity. These findings indicate a hot temperature front traveling through the high photosphere and throughout the low to mid chromosphere. This is compatible with a shock wave resulting from an oscillation induced at the location of the squeezed magnetic element. \\
\begin{figure*}
\sidecaption
\includegraphics[width=12cm]{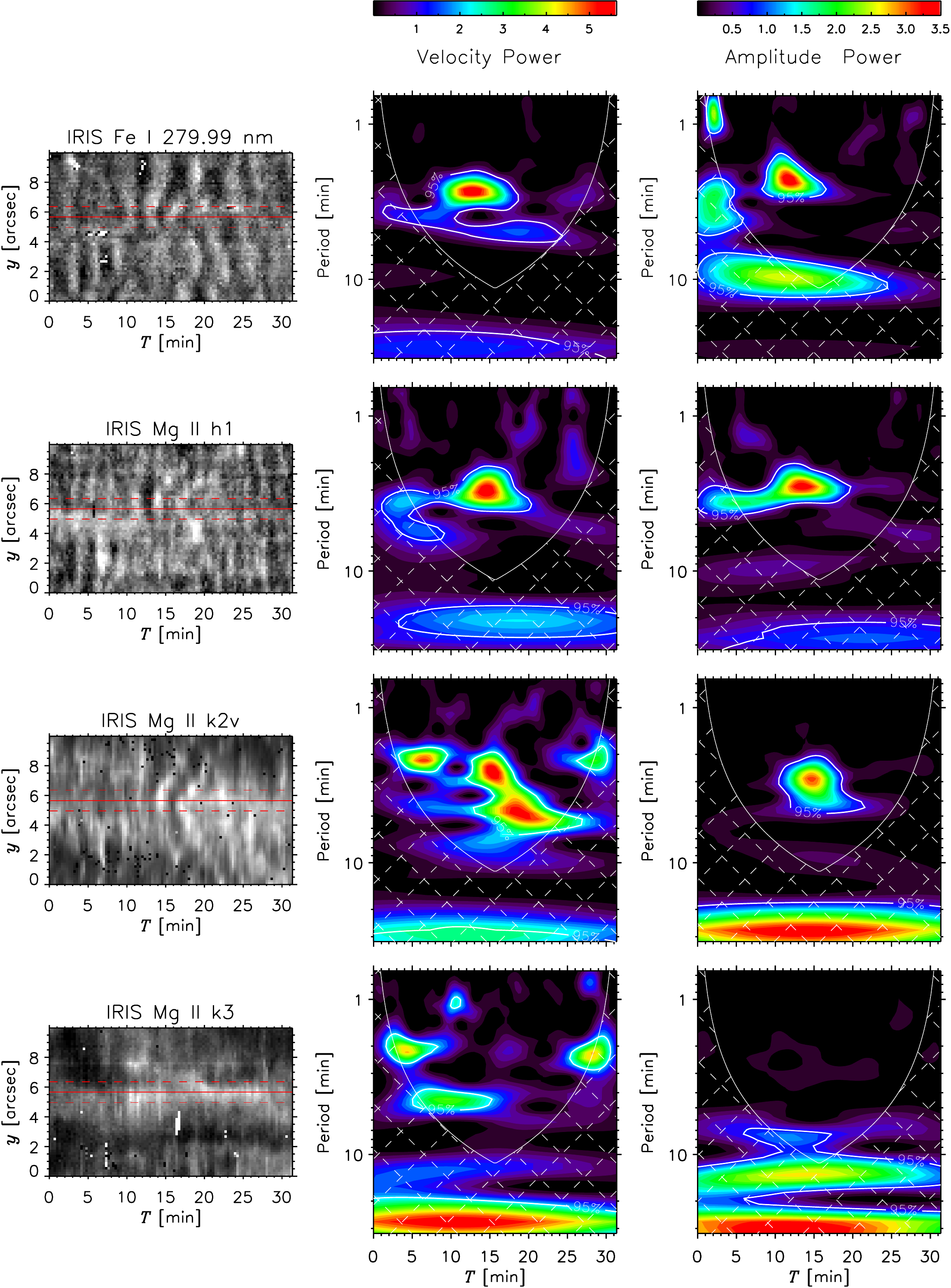}
\caption{ {\em Left to Right:} The upper row shows first a time-space image retrieved from the velocity parameter of the Fe\,{\sc i}\,2799.9\,\AA\, line ($\pm$\,$1.9$\,km\,s$^{-1}$) obtained from the  {\em IRIS} slit spectra using the MOSiC code. The dashed and solid horizontal red lines mark again the same region as in Fig.~\ref{specev}. The next two panels show the power for the velocity and the intensity fluctuations at periods ranging from 0.6\,min to $21$\,min at the slit location marked with the red solid horizontal line in the first panel. The power is shown in a logarithmic scale with the hashed area marking periods in which the wavelet analysis yields untrustworthy results. The results obtained with a confidence level of $95$\% are contoured with a white solid line (not all are labeled). The following rows are in the same format as the first row, now  showing first maps of the Mg\,{\sc ii}\,h1 minimum obtained through the MOSiC code and the Mg\,{\sc ii}\,k2v peak and the Mg\,{\sc ii}\,k3 core obtained with the ${iris\_get\_mg\_features\_lev2.pro}$ code, where blacked out or white pixels signify a failed feature finding. The power for the velocity and the intensity fluctuations for these parameters are shown in the second and third panels in each row.}
\label{isp_wave}
\end{figure*}
We apply a wavelet analysis on various parameters derived from the {\em IRIS} slit spectra using the code provided by C.~Torrence and G.~Compo ({http://atoc.colorado.edu/research/wavelets/}) using the default Morlet wavelet. 
In Fig.~\ref{isp_wave} we show the results for the velocity and intensity fluctuations in 1.) the photospheric Fe\,{\sc i}\,2799.9\,\AA\, line, 2.) the Mg\,{\sc ii}\,h1 minimum, 3.) the Mg\,{\sc ii}\,k2v peak sampling the low to mid chromosphere~\citep{2013ApJ...778..143P}, and 4.) the Mg\,{\sc ii}\,k3 sampling the high chromosphere~\citep{2013ApJ...772...90L}. 
The wavelet analysis shows a power peak in both the velocity and intensity fluctuations of the photosphere to mid-chromosphere diagnostics, with a delay increasing with height between the maximum power. The maximum power is located at a period of between 2 and 3 minutes including also larger periods in the mid chromosphere of up to $5$ to $6$\,minutes. The high chromospheric signatures, as seen with the Mg\,{\sc ii}\,k3 amplitude (rightmost panel), show no power and only a weak response in the velocity power. As seen in the left panel mapping the amplitude in intensity, there is an increase visible in the amplitude which implies a change of the transition region height~\citep{2013ApJ...778..143P}. This could signify dissipation of the shock wave energy and local heating of the high chromosphere.    
Upward-propagating shock fronts are common in quiet Sun chromosphere, both in observations \citep{2008A&A...479..213B} and numerical simulations \citep{1997ApJ...481..500C, 2004A&A...414.1121W}. We do not find a significant excess emission co-spatial with the chromospheric emission in the \ion{C}{ii}\,1335\,\AA\, line of IRIS, which is consistent with our finding that dissipation happens at the middle chromosphere.

\section{Summary and conclusion}\label{sec:disc}
A magnetic element at the edge of an exploding granule is squeezed by opposing horizontal flows, resulting in an elongation of the isocontours in magnetic flux as seen with the {\em Hinode} SP, with a chromospheric response consequently being triggered and observed in the {\em IRIS} spectra. 
Signatures of an energy deposit in the middle chromosphere are seen through a wavelet analysis of different spectral features in the
\ion{Mg}{ii}\,h \& k spectra. Our finding is consistent with an upward-propagating shock front triggered by the exploding granule.

\begin{acknowledgements}
      CEF has been funded by the DFG project Fi- 2059/1-1. NBG acknowledges financial support by the Senatsausschuss of the Leibniz-Gemeinschaft, Ref.-No. SAW-2012-KIS-5 within the  CASSDA project.
RR acknowledges financial support by the Spanish Ministry of Economy
and Competitiveness through project AYA2014-60476-P. Wolfgang Schmidt has provided  important  and  useful  comments  on  the  manuscript.
{\em Hinode} is a Japanese mission developed and launched by ISAS/JAXA, collaborating with NAOJ as a domestic partner, NASA and STFC (UK) as international partners. 
{\em IRIS} is a NASA small explorer mission developed and operated by LMSAL with mission operations executed at NASA Ames Research center and major contributions to downlink communications funded by ESA and the Norwegian Space Centre.
\end{acknowledgements}

\bibliographystyle{aa}   
\bibliography{arxiv_ms}  
%

\begin{figure*}
\centering  
\resizebox{16cm}{!}{ \includegraphics{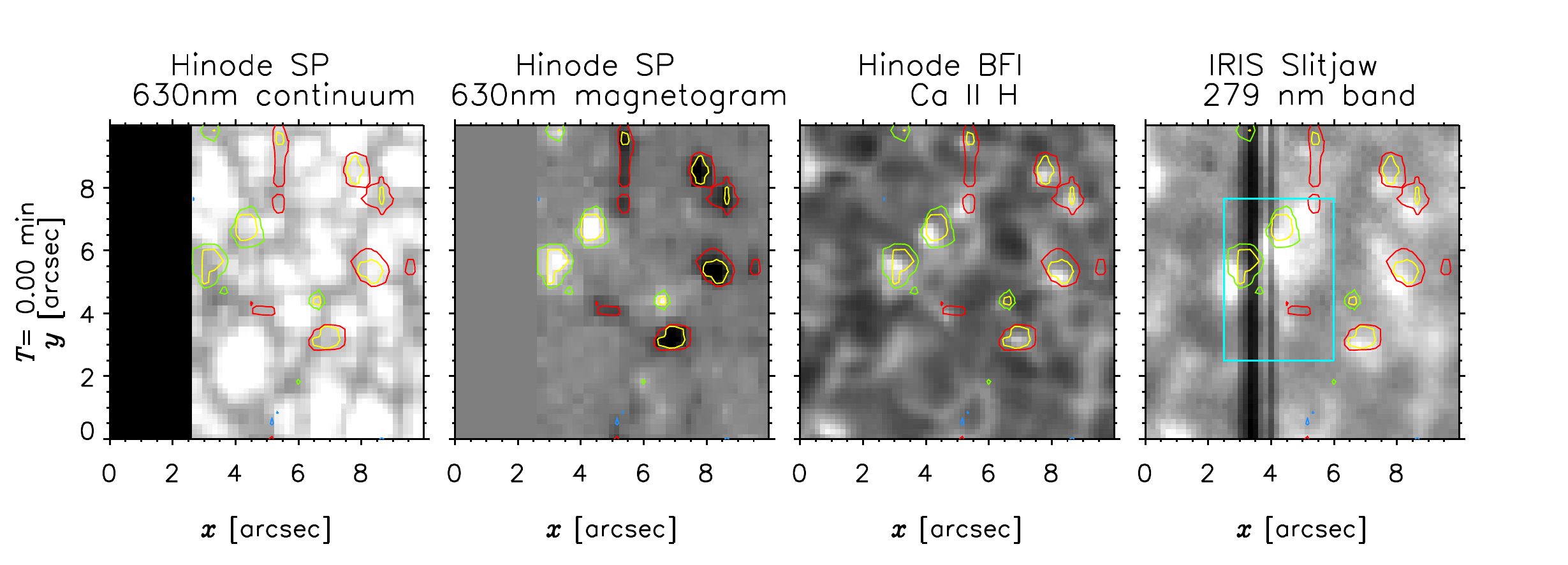}}
\caption{The first panel shows the {\em Hinode} SP 6302\,\AA\, continuum maps constructed from Stokes $I$ spectra. Part of the field-of-view is not covered by the SP slit data. The images were clipped between 10k and 16k. The following panel displays the B$_{long}^{app}$ maps clipped at  $\pm$ $80$\,Mx\,cm$^{-2}$ retrieved with the sp$\_$prep.pro code. This is followed by the {\em Hinode} Ca\,{\sc ii}\,H image (clipped to below 1500 counts) and the {\em IRIS} $2796$\,\AA\, band slitjaw image (clipped to below 200 counts). In the {\em IRIS} slitjaw images one can clearly see the {\em IRIS} slit position as a vertical dark strip. A thinner dark strip next to the slit position is due to reflections within the instrument. In all panels the contours are for $\pm$ B$_{long}^{app}$  $30$\,Mx\,cm$^{-2}$ shown with green and red colors denoting opposite polarity. The yellow contours are at levels of $60$\,Mx\,cm$^{-2}$ shown for both polarities. The contours for B$_{trans}^{app}$ $120$\,Mx\,cm$^{-2}$ are in blue. The FOV marked with a square in the last column shows the area displayed in Fig.~\ref{specev}. The times labeled on the left refer to the recording time of the {\em IRIS} slitjaw images in the $2796$\AA\,  band. For {\em Hinode} we chose the closest frames referring to this time (see also section~\ref{sec:obs} for explanation of the time frame). }
\label{Time_evolv_f}
\end{figure*}

\begin{figure*}
\centering  
\resizebox{16cm}{!}{ \includegraphics{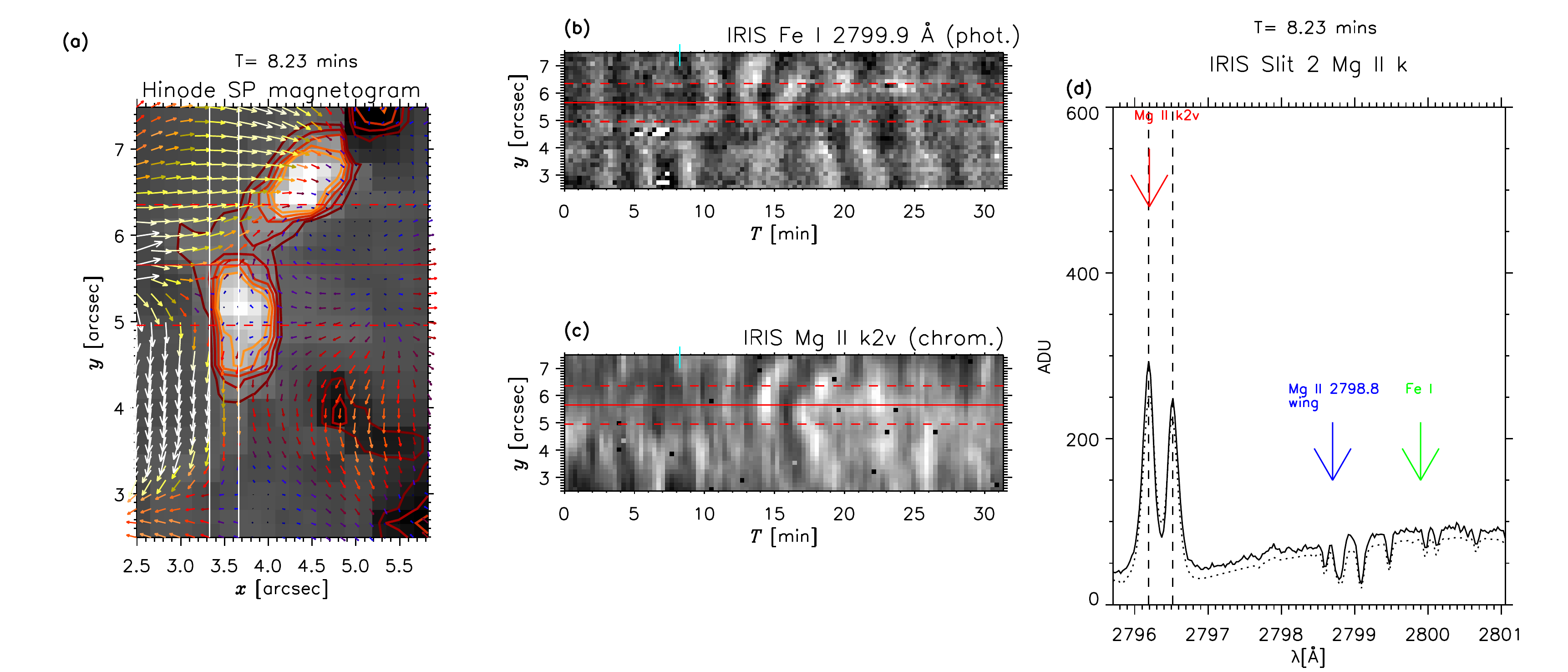}}
\caption{ Panel (a) to (c) are the same as in Fig.~\ref{specev}. Panel (a) shows the {\em Hinode} SP B$_{long}^{app}$ map now rebinned to the {\em IRIS} slitjaw scale with the same contours as in the second row of Fig.~\ref{fdev} and the horizontal flows indicated with colored arrows as in Fig.~\ref{fdev}. The two vertical white lines delimit the position of the {\em IRIS} slit in use. Two dashed red horizontal lines and a solid red line mark a region of interest. Panels (b) and (c) are time-space images derived from spectra recorded by the second {\em IRIS} slit. Panel (b) shows the line-core velocity of the Fe\,{\sc i}\,2799.9\,\AA\,  line  ($\pm\,1.9$\,km\,s$^{-1}$,  where down flows are positive) retrieved with the MOSiC code and in panel (c) the intensity of the Mg\,{\sc ii}\,k2v peak (in data units) obtained by the {\em iris$\_$get$\_$mg$\_$features$\_$lev2$.$pro} code is plotted. The red horizontal lines indicate the same area as in panel (a). The short vertical line in cyan indicates the time of the maps displayed in panel (a). In panel (d) the spectral profile recorded by {\em IRIS}  is plotted with a solid line as well as an average quiet Sun spectrum (dotted). The dashed vertical lines denote the average position of the Mg\,{\sc ii}\,k2 peaks. The arrows point to the Mg\,{\sc ii}\,k2v peak, the wing of the Mg\,{\sc ii}\,2798.8\,\AA\, line, and the position  of the  Fe\,{\sc i}\,2799.9\,\AA\,  line used in panel (b).}
\label{Spec_evolv_f}
\end{figure*}
\end{document}